\documentclass[oneside,a4paper,links]{amca}
\usepackage{hyperref}
\usepackage{graphicx}
\usepackage{amsmath,amsfonts}
\usepackage{amsmath,amssymb}
\usepackage[table]{xcolor}
\usepackage{array}
\usepackage{fancyhdr}
\usepackage[utf8]{inputenc}

\newcolumntype{+}{!{\vrule width 2pt}}

\newlength\savedwidth

\newcommand\thickhline{\noalign{\global\savedwidth\arrayrulewidth\global\arrayrulewidth 2pt}%
\hline
\noalign{\global\arrayrulewidth\savedwidth}}

\title{A Measure of Dependence Between Discrete and Continuous Variables}

\author[a,b]{Miguel A. Ré}
\author[b]{Guillermo G. Aguirre Varela}
\affil[a]{Facultad Regional Córdoba,
Universidad Tecnológica Nacional ,
Maestro López y Cruz Roja Argentina, Ciudad Universitaria, 16000~Córdoba, Argentina,
mgl.re33@gmail.com}
\affil[b]{Facultad de Matemática, Astronomía, Física y Computación,
Universidad Nacional de Córdoba,
Haya de la Torre y Medina Allende, Ciudad Universitaria, 16000~Córdoba, Argentina,
re@famaf.unc.edu.ar, guiava@gmail.com}
\begin{document}
\vspace{3cm}

\maketitle

%% To set PDF METADATA: uncomment and replace fields in
%% UPPERCASE with appropriate values. 
%% 
%% \hypersetup{
%%   pdfauthor={AUTHORS},
%%   pdfkeywords={KEYWORDS},
%%   pdftitle={TITLE}
%% }
%%
%% For instance
%% \hypersetup{
%%   pdfauthor={Sponge B. and Star P.},
%%   pdfkeywords={multiphase flow, air-liquid mixtures},
%%   pdftitle={A new model for multi-phase flow}
%% }
%%
%% NOTE: To set the metadata is recommended but not absolutely
%% neccesary. 
%% This was done before with the \pdfinfo command,
%% but according to this post:
%% http://de.nntp2http.com/comp/text/tex/2008/12/5358fd061de9703a781885a5dcf98364.html
%% if `hyperref' is used, then you must use \hypersetup{} not \pdfinfo{}


\begin{abstract}
Mutual Information (MI) is an useful tool for the recognition of mutual dependence berween data sets. Differen methods for the estimation of MI have been developed when both data sets are discrete or when both data sets are continuous. The MI estimation between a discrete data set and a continuous data set has not received so much attention. We present here a method for the estimation of MI for this last case based on the kernel density approximation. The calculation may be of interest in diverse contexts. Since MI is closely related to Jensen Shannon divergence, the method here developed is of particular interest in the problem of sequence segmentation.
\end{abstract}

\section{Introduction}
\label{sec1}
Mutual Information (MI) is a quantity theoretically based on information theory ~\cite{cover}. Since MI between two independent random variables (RV) is zero a non null value of MI between the variables gives a measure of mutual dependence.\\
When analyzing two data sets $X$ and $Y$ (assumed to be the realization of two RV) MI may give us a measure of the mutual dependence of these sets. Although MI may be straightforwardly calculated when the underlying probability distributions are known,  this is not usually the case and only the data sets are available. So MI must be estimated from the data sets themselves.
When $X$ and $Y$ are of discrete type MI may be estimated by substituting the joint probability of these variables by the relative frequency of appearance of each pair $\left( x,y\right) $ in the sequence of data ~\cite{gross,reaza}. For real valued data sets (or of discrete type with many possible values) the estimation of MI is more difficult since relative frequency or binning are not efficient methods. Alternative proposals when both data sets are of continuous type have been made ~\cite{steue}.\\
Estimation of MI between a discrete RV and a continuous one has not been so extensively considered in spite of being a problem of interestin in diverse situations. For instance we could compare the day of the week (weekday-weekend, discrete) with traffic flow (continuous) quantifying this effect. In a different context we might wish to quantify the effect of a drug (given or not, discrete) in medical treatments (epilepsia continuous data).\\
In this paper we propose a method for estimating MI between a discrete and a continuous data set based on the kernel density approximation (KDA) \cite{silve} for estimating the probability density function (PDF)  for continuous variables. For the discrete variable we make use of the usual frequency approximation~\cite{gross,reaza}. To complete the calculation of MI it is necessary to calculate an integral. This calculation is approximated by sample average.\\
MI between a discrete and a continuous RV may be identified with a weighted form of Jensen Shannon divergence (JSD)~\cite{gross}, a measure of distance between two (o more) probability distributions. JSD is of particular interest in the problem of segmentation of sequences. In this case the discrete variable is identified with the segment in the sequence and the distance between the conditional probabilities of the sequence variable is calculated.\\
Our proposal is an alternative method to that proposed by Ross ~\cite{bross} based ont the nearest neighbour method. We find our proposal more suitable for instance in the problem of sequence segmentation since it would not require a reordering of the sequence.\\
In the Methods section we present the approximations in the calculation of MI: KDA and sample average. A test of the performance of our proposal is given in the Results section by numerical experiments. In the Discussion section we consider the results obtained and the application of the method to the sequence segmentation problem.\\
\section{Methods}
\label{sec2}
We present in this section our proposal for estimating MI based on the KDA estimator of a PDF. To calculate MI we start from a sequence of data pairs $\left( x,y\right) $ where $x$ is a discrete variable and $y$ a continuous one. We assume that these data are sampled from a joint probability density $\mu \left( x,y\right) $, although unknown at first. From the joint PDF we define the marginal probabilities 
\begin{equation}
\label{e2.1}
\begin{array}{rl}
p\left( x\right) = & \int \limits _{-\infty }^{\infty }\, dy\, \mu \left( x,y\right) \\
 & \\
\phi \left( y\right) = & \sum \limits _x \mu \left( x,y\right) 
\end{array}
\end{equation}
The MI between the RVs $X$ and $Y$ is expressed in terms of these PDFs as
\begin{equation}
\label{e2.2}
I\left( X,Y\right) = \sum _x\int \limits _{-\infty }^{\infty }\, dy \, \mu \left( x,y\right) \ln \left[ \dfrac{\mu \left( x,y\right) }{p\left( x\right) \phi \left( y\right) }\right] 
\end{equation}
Note that if the variables $X$ and $Y$ are statiscally independent then $\mu \left( x,y\right) =p\left( x\right) \phi \left( y\right) $ and in this case $I\left( X,Y\right) =0$. In this way a value of $I\left( X,Y\right) \neq 0$ gives a measure of the mutual dependence of the variables. We may rewrite $I\left( X,Y\right) $ in terms of the conditional PDFs
\begin{equation}
\label{e2.3}
\mu \left( y\mid x\right) =\dfrac {\mu \left( x,y\right) }{p\left( x\right) }
\end{equation}
as
\begin{equation}
\label{e2.4}
I\left( X,Y\right) =  \sum _x\, p\left( x\right) \int \limits _{-\infty }^{\infty }\, dy \, \mu \left( y\mid x\right) \ln \left[ \dfrac{\mu \left( y\mid x\right) }{\phi \left( y\right) }\right]
\end{equation}
\subsection{Kernel density approximation}
\label{ssec-2a}
To carry out the calculation in (\ref{e2.4}) it would be necessary the knowledge of the conditional PDFs. As was already mentioned these densities are assumed unknown and they have to be estimated from the sequence itself. Here we make use of the KDA~\cite{silve}, summarized in the following. \\
Let us assume a sequence of $n$ values sampled from a PDF $f\left( y\right) $. The estimated value $f_a\left( y\right) $ at a particular value $y$ is given by
\begin{equation}
\label{e2.1.1}
f_a\left( y\right) =\frac{1}{nh}\sum _{j=1}^n \, K\left( \frac{y-y_j}{h}\right) 
\end{equation}
Here $y_j$ are the values in the sampled sequence. The kernel function $K\left( y\right) $ must only satisfy the normalization condition
\begin{equation}
\label{e2.1.2}
\int _{-\infty }^{\infty } dy \, K\left( y\right) = 1
\end{equation}
and $h$ is a smoothing parameter that avoids spurius fine structure.
In this paper we are using a Gaussian kernel with the reported optimal bandwith in ~\cite{steue}
\begin{equation}
\label{e2.1.3}
h_o\simeq 1.06\, s _m n^{-1/5}
\end{equation} 
with $s _m^2$ the variance of the sample.\\
To estimate the conditional PDFs in (\ref{e2.3})we consider separately each subset of pairs with a particular value $x$.  For each subset we approximate the conditional PDF in (\ref{e2.3}) as
\begin{equation}
\label{e2.1.4}
\mu _a\left( y\mid x\right) = \frac{1}{n_xh_o}\frac{1}{\sqrt{2\pi }}\sum _{j_x=1}^{n_x}\exp \left[ -\frac{\left( y-y_{j_x}\right) ^2}{2h_o^2}\right]
\end{equation}
Here $n_x$ is the number of data pairs with the particular value of $x$. The sum is carried over the values of y in this subset.\\
The marginal probability of $X$ is approximated by
\begin{equation}
\label{e.2.1.5}
p_a\left( x\right) = \frac{n_x}{n}
\end{equation}
and the marginal probability density of $Y$ by
\begin{equation}
\label{e2.1.6}
\phi _a\left( y\right) = \sum _x p_a\left( x\right) \mu _a\left( y\mid x\right) 
\end{equation}
We illustrate the KDA with an example: let  us consider the Gaussian joint probability distribution $\mu \left( x,y\right) $ with two possible values of $x$
\begin{equation}
\label{e2.5}
\begin{array}{rl}
\mu \left( x=1,y\right) = & \dfrac{1}{3} \dfrac{1}{\sqrt{2\pi }}\exp \left[ -\dfrac{y^2}{2} \right] \\
 & \\
\mu \left( x=2,y\right) = & \dfrac{2}{3} \dfrac{1}{\sqrt{2\pi }\sigma _g}\exp \left[ -\dfrac{\left( y-y_m\right) ^2}{2\sigma _g^2}\right] 
\end{array}
\end{equation}
with the marginal PDF for $y$
\begin{equation}
\label{e2.6}
\phi \left( y\right) =  \dfrac{1}{3} \dfrac{1}{\sqrt{2\pi }}\exp \left[ -\dfrac{y^2}{2} \right] + \dfrac{2}{3} \dfrac{1}{\sqrt{2\pi }\sigma _g}\exp \left[ -\dfrac{\left( y-y_m\right) ^2}{2\sigma _g^2}\right] 
\end{equation}
We sampled 1000 pairs from this distribution for two different values of $y_m$ and from these pairs we made an estimation of the conditional PDFs by the KDA. In figure 1 A and B we plot the probability functions in (\ref{e2.5}) and (\ref{e2.6}) for two values of $y_m$ and the corresponding approximations. 

% Place figure captions after the first paragraph in which they are cited.
\begin{figure}[!h]
\includegraphics[clip,width=1\columnwidth]{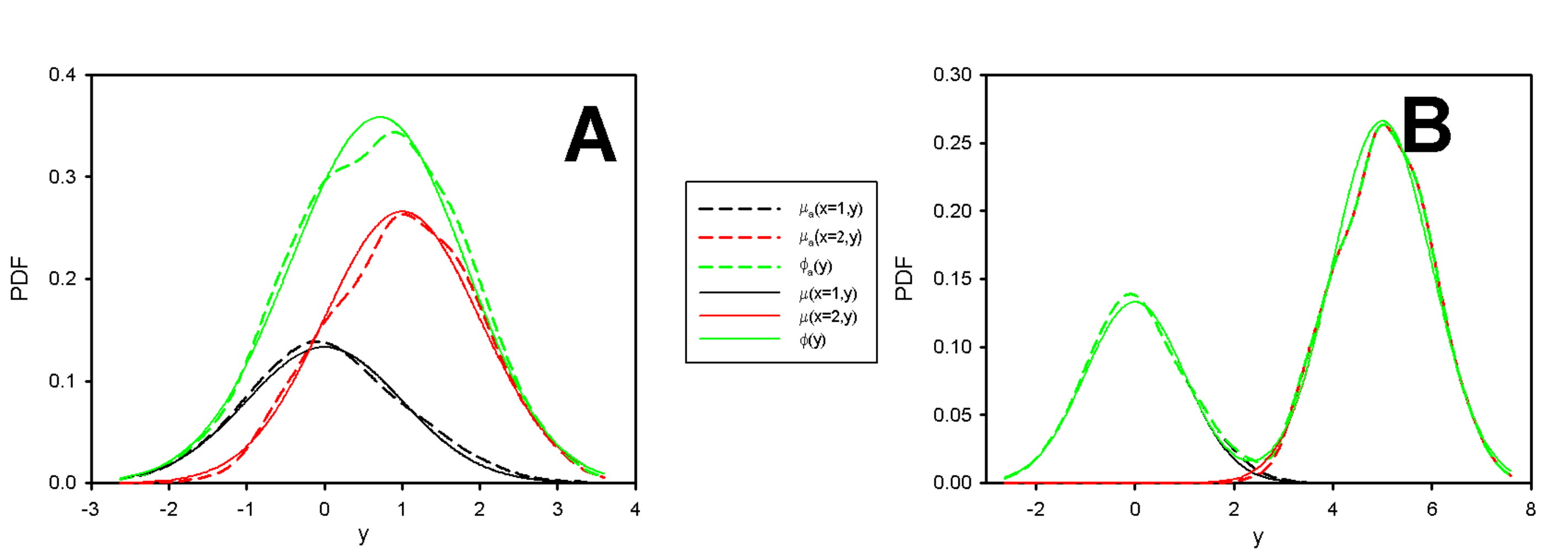}
\centering
\caption{{\bf Kernel density approximation.}
KDA for the PDFs in Eq (\ref{e2.5}) and (\ref{e2.6}) for two different values of $y_m$ and with $\sigma _g=1$. Solid lines: PDF; dashed lines KDA. A: $y_m\, =1$. B: $y_m\, = 5$.}
\label{fig1}
\end{figure}

\subsection{Sample average approximation}
\label{ssec-2b}

Habiendo aproximado las densidades de probabilidad, para completar el c\'alculo en (\ref{e2.1.4}) debemos obtener el valor de las integrales. Reconocemos en estas integrales el valor de expectaci\'on
\begin{equation}
\label{e2.2.1}
\left\langle \ln \dfrac{\mu _x\left( y\right) }{\phi \left( y\right) }\right\rangle =  \int \limits _{-\infty }^{\infty }\, dy \, \mu _x\left( y\right) \ln \left[ \dfrac{\mu _x\left( y\right) }{\phi \left( y\right) }\right]
\end{equation} 
que podemos aproximar por el promedio sobre la muestra en el segmento \citep{steue}
\begin{equation}
\label{e2.2.2}
\left\langle \ln \dfrac{\mu _x\left( y\right) }{\phi \left( y\right) }\right\rangle \simeq  \frac{1}{n_x}  \sum \limits _{j_x=1 }^{n_x}\, \ln \left[ \dfrac{\tilde{\mu }_x\left( y_{j_x}\right) }{\tilde{\phi } \left( y_{j_x}\right) }\right]
\end{equation} 
Notar que aqu\'{\i} la suma est\'a restringida a los valores en el subconjunto $x$.\\
Sustituyendo todas las aproximaciones obtenemos finalmente
\begin{equation}
\label{e2.2.3}
\tilde{D}\left[\mu _1,\mu _2\right] \simeq \frac{1}{n} \sum \limits _{x=1}^2 \sum \limits _{j_x=1 }^{n_x}\, \ln \left[ \dfrac{\tilde{\mu }_x\left( y_{j_x}\right) }{\tilde{\phi } \left( y_{j_x}\right) }\right]
\end{equation}
 \\

\section{Results}
\label{sec3}
\begin{figure}[!h]
\includegraphics[clip,width=1\columnwidth]{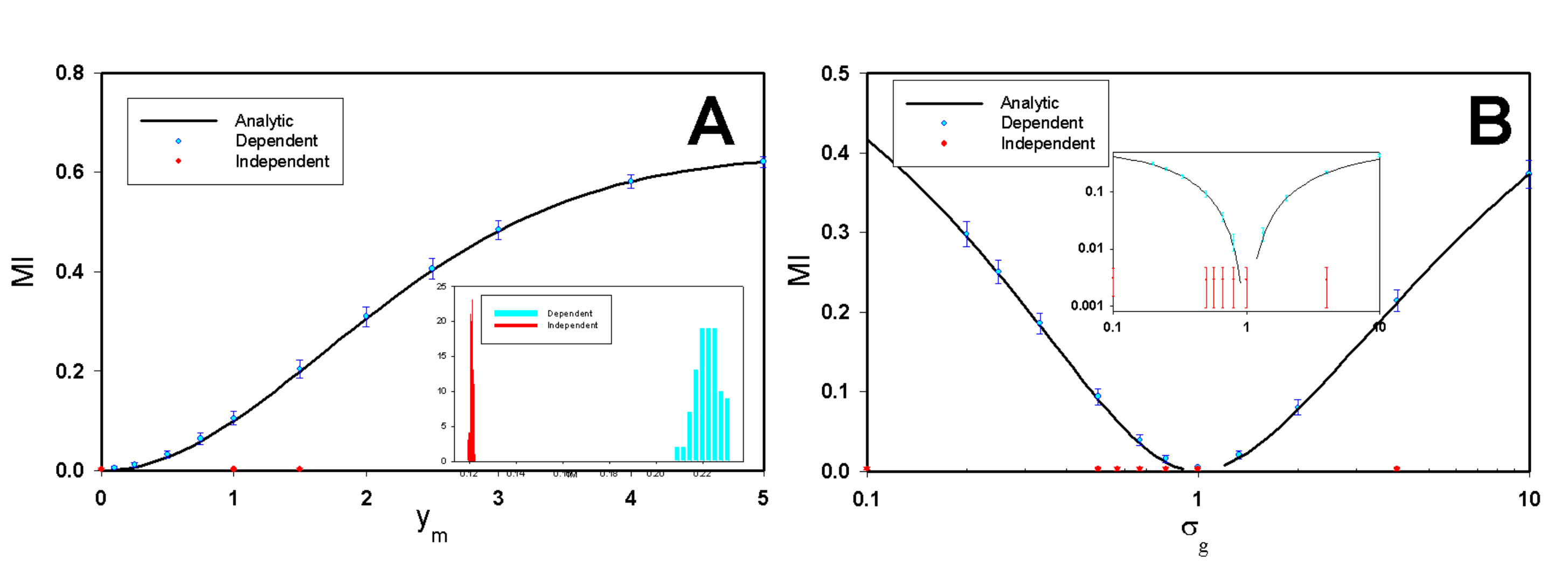}
\centering
\caption{{\bf MI estimation for Gaussian distribution.}
For the Gaussian distribution in (\ref{e2.5}) the dots represent the MI average value of 100 data sets of 1000 $\left( x,y\right) $ pairs each with the bars indicating the standard deviation of each set. The black line is the analytical value of MI. A) in function of mean value $y_m$. The inset shows the distribution of MI for a particular value of $y_m$ for a dependent and an independent set. B) changing $\sigma _g$ in the inset the same plot but in log-log scale to highlight the MI value for independent sets.}
\label{fig2}
\end{figure}

\begin{figure}[!h]
\includegraphics[clip,width=1\columnwidth]{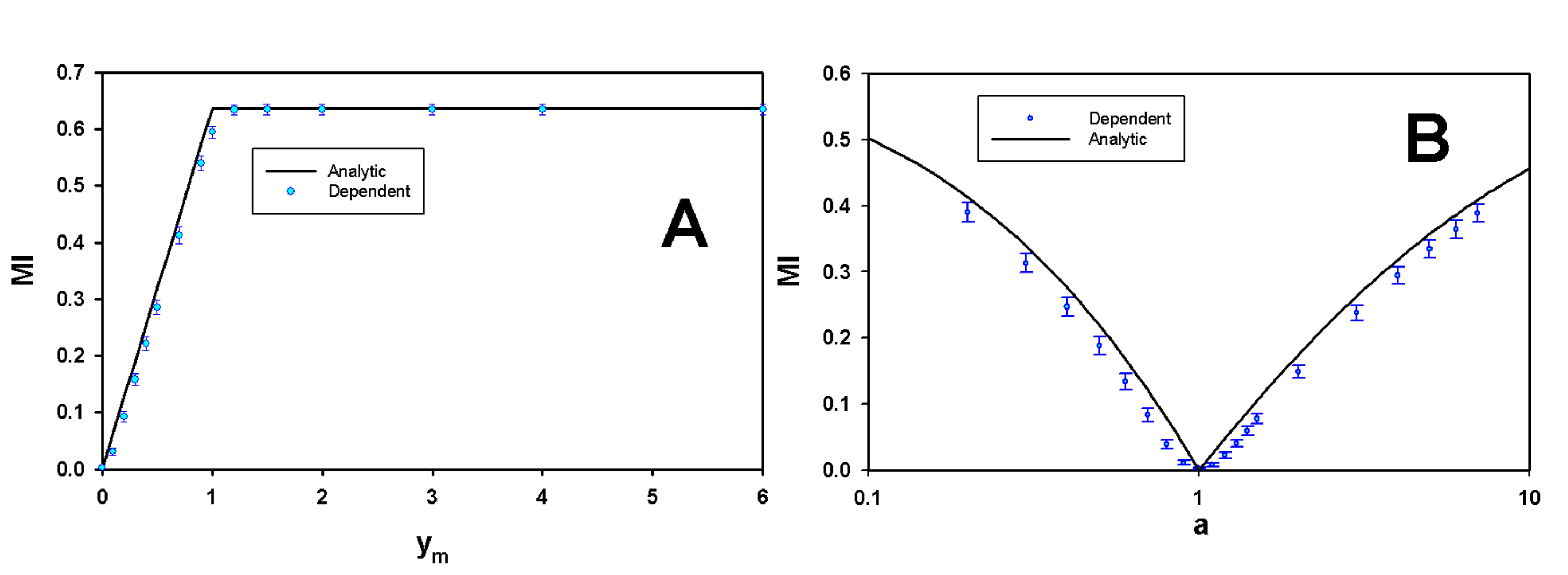}
\centering
\caption{{\bf MI estimation for uniform distribution.}
For the uniform distribution in (\ref{e3.1}) the dots represent the MI average value for 100 data sets of 1000 $\left( x,y\right) $ pairs each, with the bars indicating the standard deviation. The black line is the analytical value of MI. A) in function of mean value $y_m$. B) changing $a$.}
\label{fig3}
\end{figure}

\begin{figure}[!h]
\includegraphics[clip,width=0.5\columnwidth]{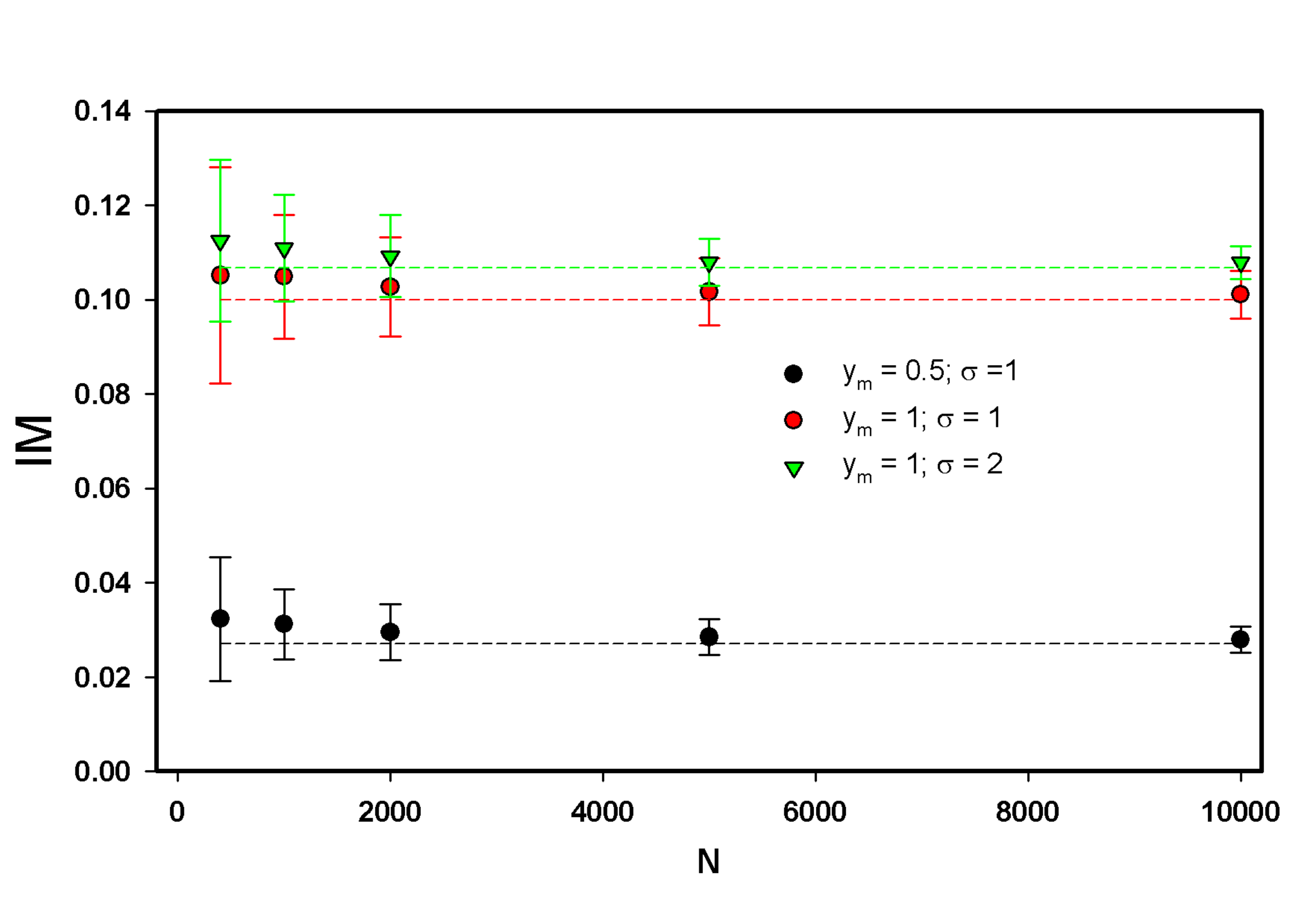}
\centering
\caption{{\bf MI estimation for Gaussian distribution.}
For the Gaussian distribution in (\ref{e2.5}) the dots represent the average value of 100 data sets for different number of $\left( x,y\right) $ pairs with the bars indicating the standard deviation. The dashed lines are the analytical values of MI for the different sets of parameters.}
\label{fig4}
\end{figure}

We test the performance of our proposed method by considering numerical experiments.\\
In the first place we consider two distributions $\mu \left( x,y\right) $: the Gaussian  distribution in (\ref{e2.5}) and an uniform distribution 
\begin{equation}
\label{e3.1}
\begin{array}{rl}
\mu \left( x=1,y\right) = & \dfrac{1}{3} \left[ \Theta \left( y+0.5\right) - \Theta \left( y - 0.5\right) \right] \\
 & \\
\mu \left( x=2,y\right) = & \dfrac{2}{3} \dfrac{1}{a} \left[ \Theta \left( y+y_m+a/2\right) - \Theta \left( y+y_m-a/2\right) \right] 
\end{array}
\end{equation}
Here $\Theta \left( y\right) $ is the step function
\begin{equation}
\label{e3.1a}
\Theta \left( y\right) = \left\{
\begin{array}{lr}
0 & \mbox{for } y<0 \\
 &  \\
1 & \mbox{for } y>0
\end{array}
\right.
\end{equation}
From each distribution we generated 100 data sets sampling 1000 $\left( x,y\right) $ data pairs with different values of $y_m$, the mean value of the distributions, or of $\sigma _g$ or $a$ respectively. We estimated the MI, $I\left( X,Y\right) $, from each set by the method described in the previous section.\\
Since we are sampling the data pairs from known distributions we are also able to calculate MI from its analytical expression. In this way we may compare the results obtained from the approximation with the corresponding analytical results.\\
These results are included in Fig. \ref{fig2} for the Gaussian distribution and in Fig \ref{fig3} for the uniform distribution respectively. We include the average value of MI over the 100 data sets for the different values of the parameters and the bars correspond to the standard deviation in each set.\\
In addition we calculated the MI for samples of statistically independent variables to establish a significance value for the MI of the dependent variables. The analytical value in this case is zero as already mentioned.\\
% Place figure captions after the first paragraph in which they are cited.
We consider the effect of the size of the sample repeating the experiment with the Gaussian distribution for different values of $n$, the number of data pairs in each set. We again generated 100 data sets of $n$ data pairs each. The results are included in Fig \ref{fig4} for three sets of parameters. As can be appreciated there is a slight overestimation of MI for small values of $n$.\\
Finally we considered an usual situation when there is only one sample of data pairs. We sampled 1000 pairs from the distributions in (\ref{e2.5}), in (\ref{e3.1}) and from the exponential distribution
\begin{equation}
\label{e3.2}
\begin{array}{rl}
\mu \left( x=1,y\right) = & \dfrac{1}{3} exp\left( -y\right) \Theta \left( y\right) \\
 & \\
\mu \left( x=2,y\right) = & \dfrac{2}{3} \dfrac{1}{2} exp\left( -y/2\right) \Theta \left( y\right) 
\end{array}
\end{equation} 
For each sample we estimated MI by the approximate method developed. These values are included in table \ref{table1}.\\
To set up a significance value for each sample we proceeded in the following way: we generated 100 data sets of 1000 pairs of independent variables, the discrete values were sampled from the distribution
\begin{equation}
\label{e3.2a}
%\begin{array}{rl} 
p\left( x\right) = \dfrac{n_x}{1000}
%\mu \left( x=2,y\right) = & \dfrac{2}{3} \dfrac{1}{2} exp\left( -y/2\right) 
%\end{array}
\end{equation}
with $n_x$ the number of times that the value $x$ appears in the original sequence and the continuous values were sampled from the Gaussian distribution
\begin{equation}
\label{e3.3}
%\begin{array}{rl} 
\mu \left( y\right) = \dfrac{1}{\sqrt{2\pi }s} exp\left[ -\dfrac{\left( y-m\right) ^2}{2s^2}\right] 
%\mu \left( x=2,y\right) = & \dfrac{2}{3} \dfrac{1}{2} exp\left( -y/2\right) 
%\end{array}
\end{equation} 
independently of the value of $x$. Here $m$ is the mean value in the original sequence and $s^2$ the variance.\\
We calculated the MI for each data set and then the MI mean value and its variance. These results are also included in table \ref{table1}. It can be appreciated a clear difference between the MI of the dependent values and those of the independent sequences.

% Place tables after the first paragraph in which they are cited.
\begin{table}[!ht] 
%\begin{adjustwidth}{-2.25in}{0in} % Comment out/remove adjustwidth environment if table fits in text column.
\centering
\caption{
{\bf MI and significance value.}}
\begin{tabular}{|l+r|l|l|}
\hline
\multicolumn{1}{|c+}{\bf PDF}  & \multicolumn{1}{c|}{\bf MI} & \multicolumn{2}{|c|}{\bf Sigificance Value}\\ \cline{3-4}
 & & mean & st. dev. \\\thickhline
{\bf Gaussian}  & 0.6359 & $4.5\times 10^{-3}$ & $1.8\times 10^{-3}$  \\ \hline
{\bf Uniform} & 0.1429 & $4.5\times 10^{-3}$ & $1.8\times 10^{-3}$ \\ \hline
{\bf Exponential} & 0.0718 & $4.5\times 10^{-3}$ & $1.9\times 10^{-3}$ \\ \hline
\end{tabular}
\begin{flushleft} MI of the sampled dependent sequences (see text) and the corresponding signifcance values computed from the independent sets.
\end{flushleft}
\label{table1}
%\end{adjustwidth}
\end{table}

\section{Discussion and Conclusions}
\label{sec4}
We have presented a method for estimating MI between a discrete and a continuous data sets. The numerical experiments described in the previous section show a good performance of this method. This can be appreciated in Figs \ref{fig2} and \ref{fig3} where the analytical results are compared with the approximate method showing a good agreement. It can be appreciated a slight underestimation of MI in Fig. \ref{fig3}. This may be attributed to the kernel selection we have made. This fact will be addressed in future research. Nevertheless approximate MI calculation is a reliable tool for dependence detection between discrete and continuous data sets.\\
The quality of the approximation depends on the size of the sample, as could be expected, being better for larger samples as is shown in Fig. \ref{fig4}. In any case the analytical value is within the standard deviation interval. This interval is reduced as $n$, the size of the sample, increases.\\
The method can also give an estimation of a significance value as described in the previous section.  This is important when deciding the dependence between the data sets.\\
MI may be identified with a weighted JSD, an entropic measure of distance between probability distributions \cite{gross}
\begin{equation}
\label{e4.1}
D=H\left[ \sum _x \pi _x \mu _x\left( y\right) \right] - \sum _x \pi _xH\left[ \mu _x\left(y \right) \right] 
\end{equation}
where, for continuous range RVs,
\begin{equation}
\label{e4.2}
H\left[ \mu _x\left( y\right) \right] = - \int dy \, \mu _x\left( y\right) \ln \left[ \mu _x\left( y\right) \right] 
\end{equation}
is Gibbs-Shannon entropy of the probability densities and $\pi _x$ denotes the weights of each density.
MI conicides with JSD identifying the weights $\pi _x$ in (\ref{e4.1}) with the marginal probability $p\left( x\right) $ in (\ref{e2.1}) and the PDFs $\mu _x\left( y\right) $ with the conditional probabilities $\mu \left( y\mid x\right) $ in (\ref{e2.3}).\\ 
JSD was used by Grosse {\it et al.} \cite{gross} to develop an algorithm to partition a nonstationary sequence of discrete RVs into stationary subsequences. With the method we have presented here for MI calculation this algorithm may be extended to the  problem of sequence segmentation of continuous RVs. The calculation does not require a reordering of the sequence as it would be necessary for instance with the nearest neighbour method \cite{bross}. Research along this line will be published elsewhere.\\

\section*{Acknowledgments}
%The heading of the Acknowledgment section and the References section must not be numbered.
%We wish to acknowledge Michael Shell and other contributors for developing and maintaining the IEEE LaTeX style files which have been used in the preparation of this template.  To see the list of contributors, please refer to the top of file IEEETran.cls in the IEEE LaTeX distribution.
The authors are grateful to SCyT - UTN for partially supporting this project through grant UTN 3559.

%References should be quoted in the text using the \emph{author-style}
%(a.k.a. \emph{Harvard style}). References can be cited in \emph{parenthetical} form \citep{DezaM,idelsohn94,meyer82,meyer82b}, or in \emph{textual} form, e.g. see \citet{idelsohn94,meyer82,meyer82b}.  References are grouped together and sorted alphabetically at the end of the article as shown in these instructions. Do not include references that are not cited in the article body. 

%If possible, internal PDF links must be generated for citations. The recommended color for links to references in the text is blue. The preferred color for links to external references, as web pages, is red (e.g. \url{http://www.amcaonline.org.ar}).

%
\bibliography{esquema}

\end{document}